# Fluctuation-Induced Interactions and the Spin Glass Transition in $Fe_2TiO_5$

P. G. LaBarre[1], D. Phelan[2], Y. Xin[3], F. Ye[4], T. Besera[3,5], T. Siegrist[3,5], S. V. Syzranov[1], S. Rosenkranz[2], A. P. Ramirez[1]

[1] *Physics Department, University of California Santa Cruz, Santa Cruz, CA 95064*
[2] *Materials Science Division, Argonne National Laboratory, Lemont, IL, 60439*
[3] *NHMFL, Florida State University, Tallahassee, FL 32310*
[4] *Neutron Scattering Division, Oak Ridge National Laboratory, Oak Ridge, TN, 37830*
[5] *Department of Chemical and Biomedical Engineering, FAMU-FSU College of Engineering, Tallahassee, FL, 32310*

ABSTRACT

We investigate the spin-glass transition in the strongly frustrated well-known compound $Fe_2TiO_5$. A remarkable feature of this transition, widely discussed in the literature, is its anisotropic properties: the transition manifests itself in the magnetic susceptibly only along one axis, despite $Fe^{3+}$ $d^5$ spins having no orbital component. We demonstrate, using neutron scattering, that below the transition temperature $T_g$ = 55 K, $Fe_2TiO_5$ develops nanoscale surfboard shaped antiferromagnetic regions in which the $Fe^{3+}$ spins are aligned perpendicular to the axis which exhibits freezing. We show that the glass transition may result from the freezing of transverse fluctuations of the magnetization of these regions and we develop a mean-field replica theory of such a transition, revealing a type of *magnetic* van der Waals effect.

Magnetic materials play a central role in condensed matter physics due to the breadth of fundamental phenomena resulting from the spin degrees of freedom and their mutual interactions [1, 2]. This allows many-body theories, in which the spin's size, dimensionality, anisotropy, and interaction range are variables, to accurately describe the properties of a wide range of different systems [3]. Of interest here is spin glass (SG), a collective state displaying a cusp in the temperature dependent susceptibility ($\chi(T)$) but lacking long range order [4]. As Anderson noted [5], SG arises from the combined effects of frustration of the antiferromagnetic (AF) interactions and quenched atomic disorder. The interplay of these factors results in a landscape of metastable states into which the spins fall and cease fluctuating in time, or freeze, below a critical "glass" temperature, $T_g$. Since the degrees of freedom are atomic spins with conserved magnitude, a signature of freezing in one direction is connected to a similar response in the other directions, a connection manifested even at the mean field level.

It is an enduring puzzle, therefore, that $Fe_2TiO_5$, in which all of the $Fe^{3+}$ spins are purely *s*-state and thus isotropic, exhibits a cusp in $\chi(T)$ *only in one direction*, as shown in Fig. 1. This feature of $Fe_2TiO_5$ has puzzled physicists over several decades and has triggered a flurry of research [6-12] including a neutron scattering study that confirmed the lack of long range order below $T_g$ = 55 K [6]. While one theory introduces anisotropy in a phenomenological way [8], the microscopic source of anisotropy, at the heart of the puzzle, has not been addressed. Here we present linear $\chi(T)$, nonlinear susceptibility $\chi_3(T)$, and neutron scattering data that show, on cooling towards $T_g$, the growth of nanoscale surfboard-shaped regions of AF-ordered spins aligned along an axis transverse to the freezing direction. We argue, using the replica formalism, that SG freezing is induced by fluctuations in the transverse magnetization between surfboard spins, the first experimental observation of a magnetic analogue to the van der Waals force.

For the present study we used single crystals of $Fe_2TiO_5$ grown by J. P. Remeika. The pseudobrookite structure of $A_2BO_5$ with space group *Cmcm* was confirmed by X-ray diffraction. We note that the $Fe^{3+}$ and $Ti^{4+}$ ions share the A and B sites: the A site contains 0.64 Fe and 0.36 Ti while the B site contains 0.72 Fe and 0.28 Ti. To confirm that all the Fe ions are in the isotropic $Fe^{3+}$ state, electron energy loss spectroscopy (EELS – see supplemental material) was used. We find both L3 and L2 spectra that are fully consistent with only $Fe^{3+}$. Thus, the concentration of magnetic ions ($Fe^{3+}$) can be taken as 2/3 on both A and B-sites with random site occupation.

Magnetization measurements were performed in two different Magnetic Property Measurement Systems (MPMSs). For $\chi(T)$ and $\chi_{nl}(T)$ for $T < 300$ K, a conventional sample holder was used. For measurements with 300 K $< T <$ 900 K, the sample was mounted with Zircar cement on a rod designed for high-temperature studies. The leading non-linear susceptibility ($\chi_{nl}(T)$) was obtained by measuring magnetization ($M$) as a function of magnetic field ($H$) at fixed $T$. The $M(H)$ data were fit to a sum of polynomials ($\chi_0 + \chi_1 H + \chi_3 H^3$). Neutron scattering measurements were performed on the *Corelli* instrument [13] at the Spallation Neutron Source at Oak Ridge National Laboratory from 5 K – 300 K. *Corelli* employs a broad band of incident neutron energies to perform time-of-flight Laue diffraction measurements and simultaneously provides both the energy integrated signal, *e.g.* the equal-time two-particle correlation function, as well as the purely elastic signal, through implementing cross-correlation with a pseudorandom chopper [14]. Measured raw neutron data were transformed and put into uniform sized bins in momentum transfer space (*h,k,l*) using the Mantid software package [15, 16].

The inverse susceptibility from 5 K to 900 K is shown in Fig. 1. We find an effective moment ($\mu_{\text{eff}}$) for $H||c$ and $H \perp c$ to be 6.07 $\mu_B$ and 6.18 $\mu_B$ respectively, in reasonable agreement with an s-state moment of S = 5/2 ($\mu_{\text{eff}}$ =5.92 $\mu_B$) and a g-factor of 2.06. The antiferromagnetic (AF) Weiss constant ($\theta_W$) is found to be 893 K and 948 K for $H||c$ and $H \perp c$. Since the $Fe^{3+}$ occupation is 2/3 of all A and B sites, these $\theta_W$ values represent 2/3 that of the structure fully occupied with $Fe^{3+}$, which is not attainable. In Fig. 2 are shown $\chi_1(T)$ and $\chi_3(T)$. The behavior of $\chi_1(T)$ is similar to that previously reported [6], exhibiting SG freezing below $T_g$ K for $H||c$, with no anomaly in the other two directions. This conclusion is supported by measurements of $\chi_3(T)$ for $H \parallel c$ where a peak is seen at $T_g$. While there is only a weak decrease in $\chi_3(T)$ below $T_g$, the high temperature behavior obeys the form $\chi_3(T) = t^{-\gamma}$, Fig. 2 (inset), where $t = |T - T_g|/T_g$ and the critical exponent $\gamma = 2.72 \pm 0.09$. This value of $\gamma$ is larger than for canonical SGs, which exhibit $\gamma \approx 2.1 - 2.3$ [4], a result that is consistent with the neutron-determined correlation lengths discussed below. It is not unusual for SGs to develop Ising character, but the situation in Fe₂TiO₅ is qualitatively different since the response for $H \perp c$ is completely independent of the singularity observed for $H \parallel c$. As alluded to above, this behavior is incompatible with a mean field description of a SG based on atomic spins, and suggests that the degree of freedom itself emerges from a collective effect.

To probe collective effects amongst the atomic spins, we performed neutron scattering measurements. The upper panels of figures 3(a) and 3(b) show the elastic neutron scattering intensity measured at $T = 5$ K in the ($hk0$) and ($\frac{1}{2}kl$) planes, respectively. Here, we utilize relative reciprocal lattice units for the wavevector transfer $\mathbf{Q} = (h \times 2\pi/a, k \times 2\pi/b, l \times 2\pi/c)$, with orthorhombic lattice constants $a = 3.732$ Å, $b = 9.8125$ Å, and $c = 10.0744$ Å. Streaks of scattering along $\mathbf{b^*}$ are seen both in the ($hk0$) plane at half-integer values of $h$, as well as in the ($\frac{1}{2}kl$) plane at integer values of $l$. The presence of these streaks, well within the SG state, indicates that the correlation length ($\zeta_b$) along $\mathbf{b}$ remains short, limited to one or two nearest neighbors, while $\zeta_a$ and $\zeta_c$ both have grown to at least an order of magnitude larger than $\zeta_b$. While the pseudobrookite structure gives rise to a rather complex spin lattice, nearest spins are simply separated by a lattice unit $a$ along $\mathbf{a}$. The occurrence of narrow peaks along $\mathbf{a^*}$ at half-integer values of $h$ therefore immediately reveals the presence of strong AF correlations between neighboring spins along $\mathbf{a}$. Furthermore, cuts through these diffuse streaks of scattering along $\mathbf{a^*}$ or $\mathbf{c^*}$ are well represented by resolution-broadened Lorentzian line shapes, from which we extract correlation lengths of $\zeta_a = 42$ Å and $\zeta_c = 12$ Å at $T = 5$ K. Cuts along $\mathbf{b^*}$ (see Fig. 3c) however are very broad, do not peak at integer or half integer reciprocal lattice positions, and are poorly represented by Lorentzian profiles. This clearly indicates that the correlations are limited to a few nearest spins only along $\mathbf{b}$.

These observations allow us to develop a model of the spin arrangement as follows. First, we assume that both A and B sites are fully occupied by $Fe^{3+}$ spins. We find that the $\mathbf{b^*}$-dependence is well reproduced when correlations along the $\mathbf{b}$ are limited to nearest neighbors within the same $\mathbf{a}$-$\mathbf{c}$ plane in the form of corrugated double chains, as indicated in Fig. 4a. Including correlations that extend further along $\mathbf{b}$ to spins between different $\mathbf{a}$-$\mathbf{c}$ planes, e.g. spins in the $y=0$ and $y=\frac{1}{2}$ planes, immediately leads to much sharper peaks than observed, showing that the development of spin correlations along $\mathbf{b}$ is strongly suppressed. A likely scenario is the complete suppression of the exchange fields from spins in the $y=0$ and $y=1$ planes seen by a spin in the $y = 0.5$ plane due to AF alignment along $\mathbf{a}$ and the centering of the *Cmcm* spacegroup (see Fig. 4a). The direction of spin alignment can be inferred by observing that in the ($hk0$)-plane (Fig. 3a), the neutron scattering intensity vanishes along $\mathbf{a^*}$, e.g. for wavevectors $\mathbf{Q} = (h,0,0)$. This indicates that the spins point parallel or antiparallel to *a*, since only the spin component perpendicular to the wavevector contributes to the magnetic neutron scattering cross section [17]. In contrast, if the spins were pointing along *b*, the intensity would be maximized along **Q** = (*h*,0,0) but strongly suppressed at **Q** = (0.5,*k*,0), which is clearly not the case (see Figs. 3a and 3b, respectively). Based on this model, we calculate the neutron scattering intensity according to:

$$\frac{\partial \sigma}{\partial \Omega} \sim \sum_{\alpha,\beta} \left(\delta_{\alpha\beta} - \hat{Q}_\alpha \hat{Q}_\beta\right) f(Q)^2 \sum_{\substack{i,j \\ i \neq j}} \langle S_i \cdot S_j \rangle e^{-i\mathbf{Q}\cdot(\mathbf{R}_j - \mathbf{R}_i)}, \quad (1)$$

where *f(Q)* is the magnetic form factor for $Fe^{3+}$, $\alpha, \beta = x, y, z$, and $S_i$ the spin at position $R_j$. Here, we approximate the decay of the spin correlations along *a* and *c* in the Ornstein-Zernike form utilizing the correlation lengths $\zeta_a$ and $\zeta_c$ obtained from the Lorentzian fits to the data [18]:

$$\langle S_i \cdot S_j \rangle = S_i \cdot S_j \frac{e^{-\left(\frac{|x_j - x_i|}{\zeta_a} + \frac{|z_j - z_i|}{\zeta_c}\right)}}{|\mathbf{R}_j - \mathbf{R}_i|} \quad (2)$$

The simulated neutron intensity is then obtained by summing Eq. (1) over a box that is much larger than $\zeta_a$ and $\zeta_c$ along *a* and *c*, respectively, but very narrow along *b*, spanning only half a unit cell (e.g. only one double chain along *b*) as required in order to reproduce the *k*-dependence of the observed neutron scattering intensity. The simulated neutron intensity so obtained is in overall good qualitative agreement with the observation (see lower panels of Fig. 3a and Fig 3b). We then repeat the calculation with a random distribution of 33% of each site with spin $S_i$ = 0, close to the distribution of (Fe,Ti) in $Fe_2TiO_5$. The resulting simulated intensity (lower right panels of Fig. 3a and Fig. 3b) is again in good qualitative agreement with the observation, and qualitatively not different from the results in which all sites are fully occupied with $Fe^{3+}$ spins.

Figure 4 (c) shows the temperature dependence of both the *elastic* and the energy-integrated (total) scattering intensity integrated over one of the broad peaks along *k* at **Q** = (0.5, 3.3-5.2, 0). At high temperature, the total scattering is much stronger than the elastic signal, reflecting the presence of fluctuating spin-correlations and an elastic signal only present because of the quasielastic nature of the fluctuations. Upon cooling, the elastic signal increases faster than the total signal, with a marked sharp increase below 100 K and an inflection point near $T_g$, signaling the onset of spin freezing. However, the elastic signal remains weaker than the total signal down to 5 K, indicating that spin fluctuations are still present at base temperature. The correlation lengths along *a* and *c* further show the rapid growth of spin correlations along *a* upon cooling whereas along *c*, substantial correlations only become apparent below 150 K, see Fig 4 (b).

The neutron measurements describe a situation shown schematically in Fig. 4 (b), namely a low-temperature state of elongated, surfboard-shaped correlated regions in which the spins are aligned (anti)parallel to *a*. While the fraction of $Fe^{3+}$ spins occupying these regions is not known, the continuing increase of $\chi(T)$ along *a* and *b* on cooling below $T_g$ suggests that a sizeable fraction of spins are located in between surfboards, a region we will refer to as the *cloud*. Assuming that all of the anisotropy in spin space is associated with surfboard formation, a natural source of such anisotropy is the spin orientation in the surfboards. As is well known, $\chi(T)$ of a Heisenberg antiferromagnet becomes anisotropic below the Néel temperature $T_N$ – along the ordered-spin direction, $\chi(T)$ approaches zero as $T \to 0$, whereas transverse to this direction, $\chi(T)$ remains finite. Absent crystal field anisotropy, the spin direction is fixed to the crystal structure by the dipole-dipole energy, usually very small – e.g. in $MnF_2$ this energy is only a few K whereas $T_N$ = 67 K [19]. Here, the intra-surfboard spins order with an AF pattern in registry with the lattice, along *a*, but should retain a near-constant susceptibility transverse to *a*, suggesting a scenario in which the *c*-direction fluctuations in magnetization induce an interaction with other surfboards – a type of *magnetic* van der Waals interaction.

In order to assess the feasibility that such a van der Waals-type interaction can lead to SG freezing, we consider a model of the surfboard free energy, given by:

$$F(\mathbf{M}) = \sum_i g(M_i) - \frac{1}{2}\sum_{i,j} J_{ij} M_i M_j. \tag{1}$$

where the indices $i$ and $j$ label the surfboards, $g(M)$ is a large positive definite function with a minimum at $M = 0$; the interaction $J_{ij} = J(\mathbf{r}_i - \mathbf{r}_j)$ between the surfboards at random positions $\mathbf{r}$ is weak and vanishes when averaging over the direction of the vector $\mathbf{r}_i - \mathbf{r}_j$, which is obeyed, for example, by dipole-dipole interactions [20].

To study the glass transition, we consider the disorder-averaged partition function of the system in the replica representation [21] in the form

$$Z = \int \mathcal{D}M \, exp\left(-\frac{1}{T}\left[\sum_{i,\alpha} g(M_i^\alpha) - \frac{1}{4V}\sum_{i,j \neq i, \alpha, \gamma} J_2 M_i^\alpha M_i^\gamma M_j^\alpha M_j^\gamma\right]\right), \tag{2}$$

where $\alpha$ and $\gamma$ is the replica index and $n$ and $J_2 = \int J(r)^2 dr$. The replicated partition function (2) may be derived from the free energy (1) in the mean-field approximation, i.e. assuming that products of magnetizations fluctuate weakly on top of their average values. Going beyond the mean-field approximation will only shift the phase boundary of the glass phase by a factor of order unity.

We develop a mean-field theory of the transition [20] with the glass order parameter [21, 22] $\mathbb{Q}^{\alpha\gamma} = \frac{1}{V}\sum_i \langle M_i^\alpha M_i^\gamma \rangle$. In the replica-symmetric approximation ($\mathbb{Q}^{\alpha\gamma} = q$ for all $\alpha \neq \gamma$), the free energy of the system of surfboards is given by

$$F(q) = q^2 \frac{VJ_2(T)}{4T}\left[\frac{J_2(T)n_0(T)}{T^2}\langle M^2\rangle^2 - 1\right] + O(q^4), \quad (3)$$

where $\langle M^2 \rangle$ is the average square of the fluctuation of the magnetization $M$ of one surfboard and $n_0$ is the spatial density of the surfboard. The free energy (3) signals a phase transition at $(J_2 n_0/T^2)\langle M^2\rangle^2 = 1$.

We can write $\chi(T) = n_0 \langle M^2 \rangle/T$, and the condition for the transition becomes $1 = (\chi^2/n_0)\int J(r)^2 dr$. If we assume also that the interactions between the surfboards are dipole-dipole interactions [$J(r) \sim 1/(\mu r^3)$], the transition occurs at

$$\mu^2 n_0 r_0^3 \sim \chi^2, \quad (4)$$

where $r_0$ is the average cluster size and $\mu$ is the magnetic permeability. We consider $\chi$ in Eq. 4 to be the difference between $\mathbf{H} \parallel \mathbf{c}$ and $\mathbf{H} \perp \mathbf{c}$ at $T_g$, namely the surfboard contribution to the susceptibility, which yields a value of $1.7 \times 10^{-4}$ in dimensionless units. Thus, for $\mu = 1$, appropriate for a cloud spin medium with a similarly small susceptibility, this implies that the ratio of the average surfboard size to the average separation between surfboards is $n_0^{1/3} r_0 = 7.0 \times 10^{-3}$ = 1/320. This estimated value of $n_0^{1/3} r_0$ implies an extremely long-range interaction among surfboards but such a large disparity between the size and separation of surfboards is likely a result of the mean field approximation used to derive Eq. 4. Specifically, in 3D it is known that mean field treatments can overestimate the freezing temperature by more than a factor of two [23, 24]. In addition, the absence of correlation length growth along $\mathbf{b}$ suggests that the effective dimensionality of our system is between 3D and 2D, the latter of which is thought to exhibit a SG transition only at $T = 0$. Thus, a more exact confrontation of our model with experimental data will likely come from computational studies. While higher Fe density is not possible without compromising charge neutrality, compounds with lower Fe density will be studied in order to further test the theoretical prediction in Eq. 4.

In conclusion, we have shown that the anomalous spin glass freezing in $Fe_2TiO_5$ is associated with the growth of nano-scale surfboard-shaped regions of ordered spins. Lacking a well-defined moment themselves, the surfboards will possess strong transverse spin fluctuations that we have shown, using a mean field formalism, can lead to spin glass freezing in only one direction. This is the first experimental example of a purely magnetic analogue of the van der Waals interaction, but such an interaction might be relevant in other systems of nano-scale magnets such as possible *anti*-ferro fluids and artificial spin ice [25]. Furthermore, the well-known connection between the van der Waals force and the Casimir effect [26], in conjunction with the AF version of the former as discussed here, suggests a possible attractive force between plates of AF-ordered spins lying in the plane of the plate. Experiments to measure this force may be possible using micro-electro mechanical systems.

Acknowledgments – We acknowledge useful discussions with A. P. Young, W. P. Wolf, D. Huse, and G. Aeppli. This work was supported by the U.S. Department of Energy grant DE-SC0017862 (P.G.L, A.P.R.) and DE-SC0008832 (T.B. and T.S.). Work by Y.X., T.B. and T.S. was carried out in part at the National High Magnetic Field Laboratory, which is funded by the National Science Foundation under grant NSF-1644779 and the State of Florida. Work at Argonne (S.R., D.P., neutron scattering and high temperature susceptibility measurements) was supported by the U.S. DOE, Office of Basic Energy Science, Materials Science and Engineering Division. Use of the Spallation Neutron Source at ORNL was sponsored by the Scientific User Facilities Division, Office of Basic Energy Sciences, U.S. Department of Energy.

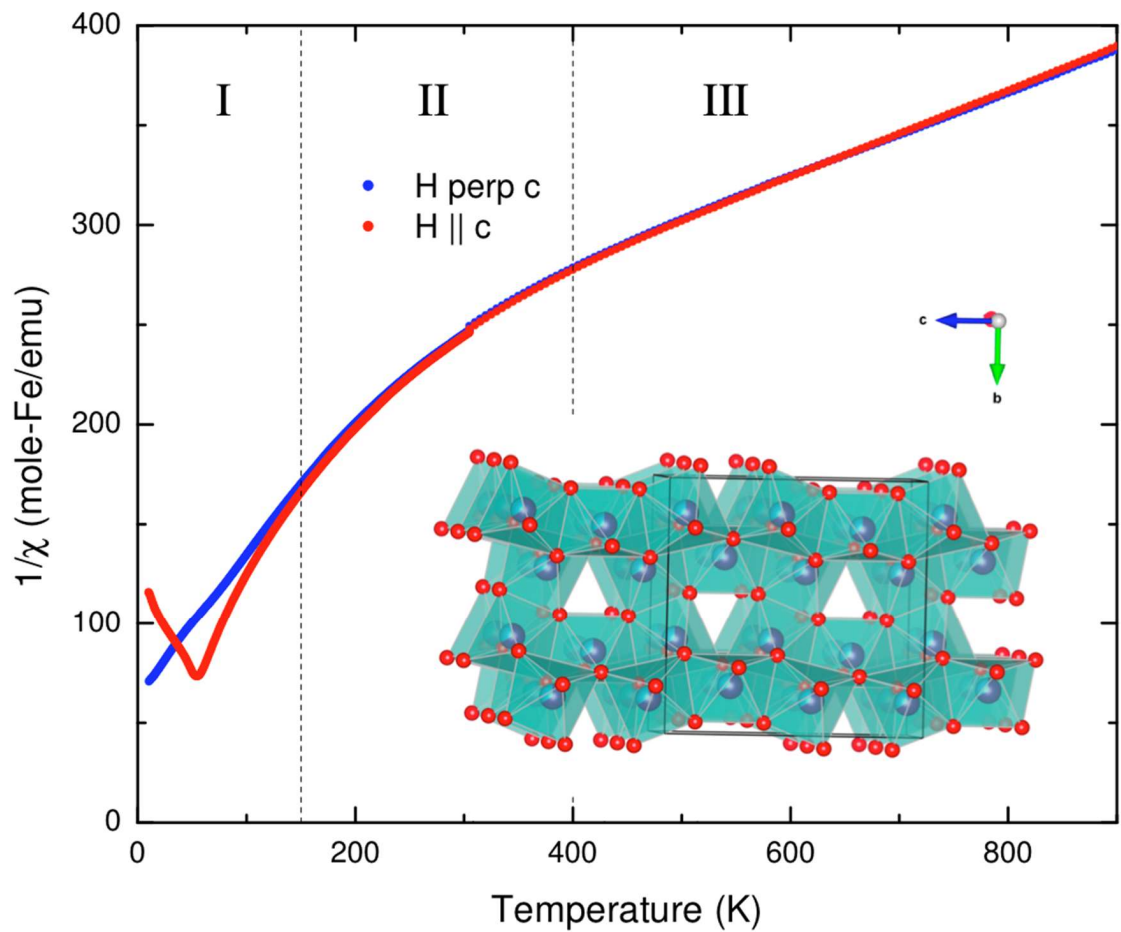

Fig. 1. Inverse susceptibility of $Fe_2TiO_5$ for H perpendicular and parallel to the c-axis. Inset: structure of $Fe_2TiO_5$ showing the Fe/Ti sites (blue balls) and oxygen (red balls).

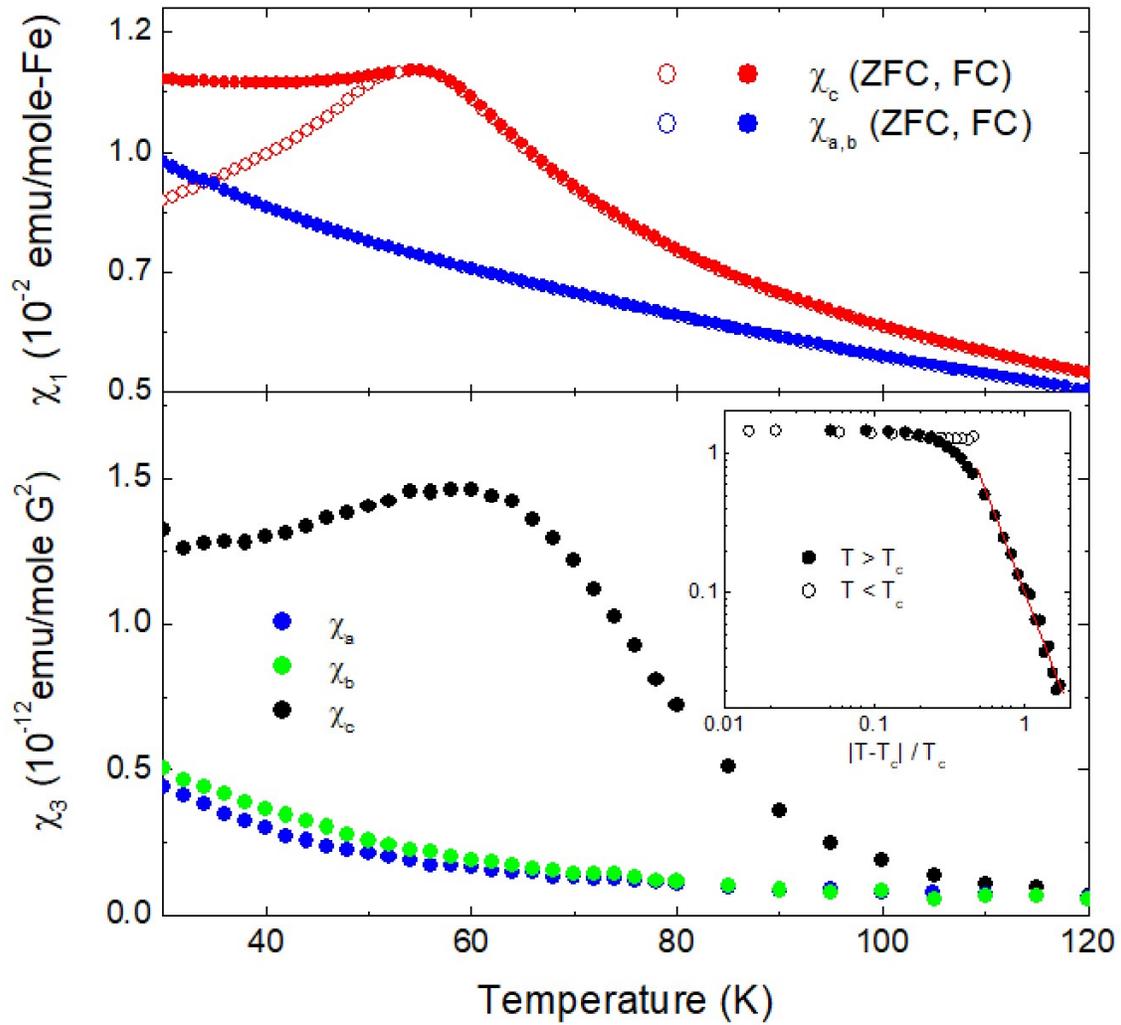

Fig. 2 Top: Linear susceptibility ($\chi_1$) in the *c* and *a*, *b* directions, for cooling in zero field (ZFC) and cooling in a field (FC), of $H = 0.1$ T. Bottom: Non-linear susceptibility ($\chi_3$) for the *a*, *b*, and *c* directions showing a peak at the freezing temperature, $T_g$, in the *c*-direction. The inset shows the power law behavior critical behavior above $T_g$ and the straight line corresponds to $\gamma = 2.72$, as discussed in the text.

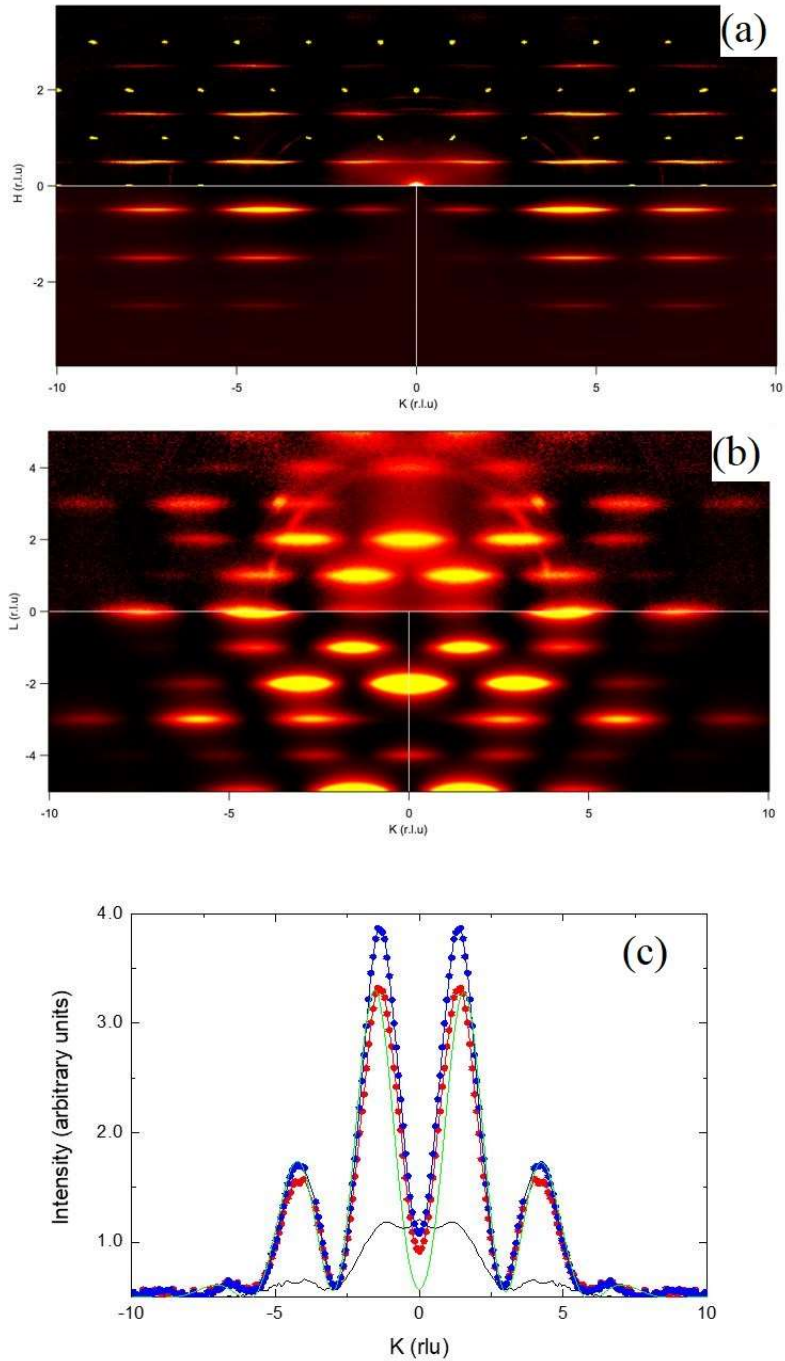

Fig. 3 Elastic neutron scattering intensity in the (a) [$h,k$,0] and (b) [0.5, $k$, $l$] planes at T=5K. The upper panels show the measured intensities with Bragg peaks visible in (a). The lower left panels show the intensities for the simulations as described in the text for the structure with both A and B sites fully occupied by $Fe^{3+}$ spins (left panels) and for a random occupation of 1/3 of the positions with $s$=0 (lower right panels) (c) Neutron scattering intensity along the $k$-direction for Q = (0.5, $k$ , 1) showing the purely elastic (red dots) as well as total scattering (blue squares) at T=6K and the much weaker total scattering at 300K. The green solid line is the calculated intensity as described in the text.

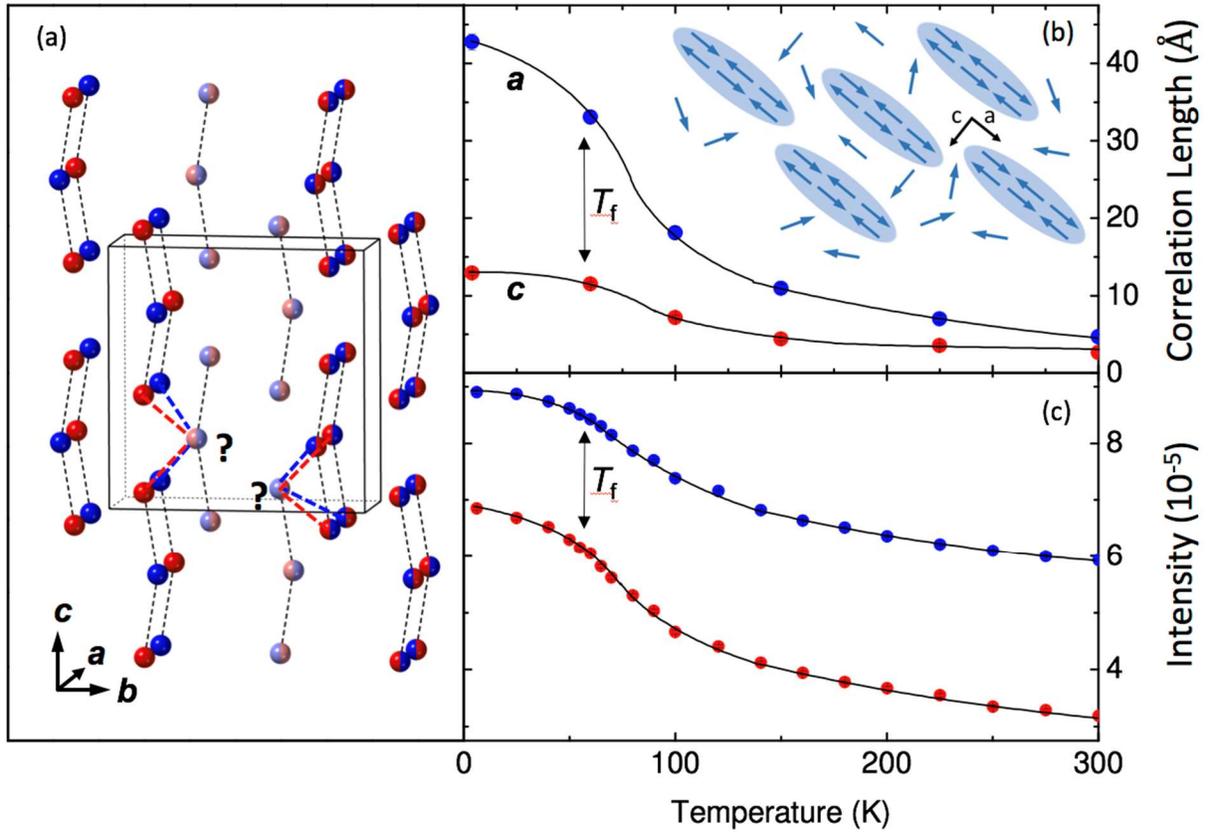

Fig. 4 (a) Illustration of the low temperature spin order in $Fe_2TiO_5$. Only the Fe/Ti positions are shown, assuming that all positions are fully occupied with Fe spins with blue (red) spheres denoting the up (down) positions along the a-direction. The spheres at right are half-colored denoting the lack of LRO. The spins are AF-aligned within the triplets indicated with light dashed lines, forming corrugated double chains (in planes at integer x positions). Along the b-direction, the next such double chain is centered at half-integer x-positions, and since the spins are AF-aligned along the a-axis, each spin at the half-integer positions sees a net zero mean field. (b) Correlation lengths in the *a* and *c* directions vs. temperature. (c) Temperature dependence of total (blue) and elastic (red) neutron scattering intensity over the wavevector range Q=(0.5±0.1, 4.25±0.95,0±0.1).

Supplemental Information for
"Fluctuation-Induced Interactions and the Spin Glass Transition in $Fe_2TiO_5$"

The Fe valence was investigated by electron energy loss spectroscopy (EELS) in a cold cathode field emission JEOL JEM-ARM200cF transmission electron microscope using a GIF QuatumSE system. The Fe $L_{2,3}$ core loss spectra were taken from freshly crushed crystal pieces in the diffraction mode. Each spectrum was collected with an acquisition time of 2s at an energy dispersion of 0.1 ev/channel. The final spectrum, shown in Fig. S1, is a sum of five spectra, which have been aligned, plural scattering removed, and background subtracted.

It has been shown that the Fe valence and coordination can be determined by the characteristic energy loss near edge structure (ELNES) [1,2]. The Fe core loss probes the 3d empty states, showing L3 (transitions from 2p3/2 to empty 3d orbital) and L2 (transitions from 2p1/2 to empty 3d orbital). Although $Fe^{2+}$ can be distinguished from $Fe^{3+}$ from the onset energy of $L_3$ peak, it is more accurate and convenient to distinguish $L_{2,3}$ ELNES since the experimental conditions affect the precise peak position. It has been clearly revealed in the previous report that ferrous $Fe^{2+}$ has a small peak on the higher energy side of the $L_3$ major peak, while it is just the opposite that ferric $Fe^{3+}$ has a small peak on the lower energy side of the $L_3$ peak. In addition, the $L_2$ peak ELNES is different as well. For $Fe^{2+}$ ELNES, the first peak (lower energy side) is much higher than the second peak, while for $Fe^{3+}$, both peaks have almost the same height [2]. Therefore, as shown in Fig. S1, the Fe $L_{2,3}$ core loss from $Fe_2TiO_5$ has a small lower energy side peak for $L_3$ and two equal height peaks for $L_2$, consistent with Fe3+.

References:
[1] L. a. J. Garvie and P. R. Buseck, Nature, 396 (1998) 667-670.
[2] A. P. Brown, S. Hillier and R. M. D. Brydson, IOP Conf. Series: Journal of Physics Conf. Series 902 (2017) 012016.

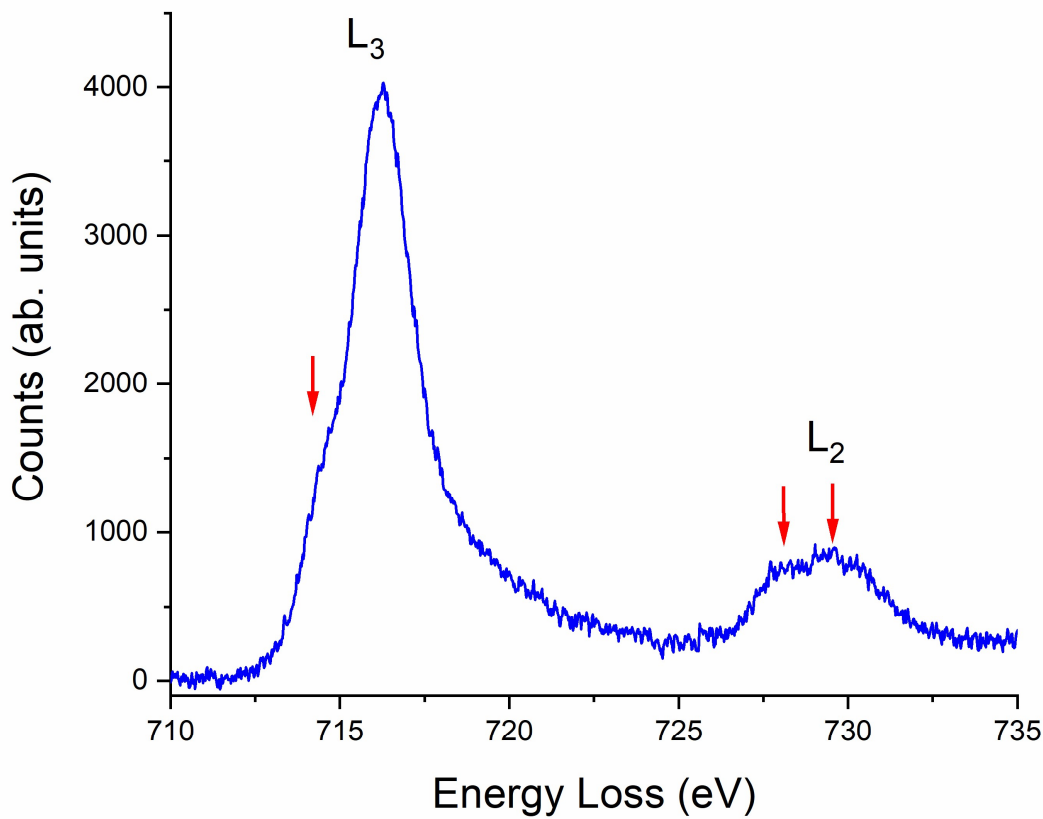

Fig. S1. Electron energy loss spectrum for $Fe_2TiO_5$.

Appendix A
Theoretical Considerations

Here we describe a model of clusters of antiferromagnetically ordered spins, depicted schematically below in Fig. A1. The average magnetization of each cluster is zero but may fluctuate. As a result of these fluctuations, the spin clusters may weakly interact with each other. We assume for simplicity that all magnetizations fluctuate along one direction, consistent with the surfboard picture in $Fe_2TiO_5$. We further demonstrate that the system exhibits a glass phase transition and estimate the phase boundary of the glass phase.

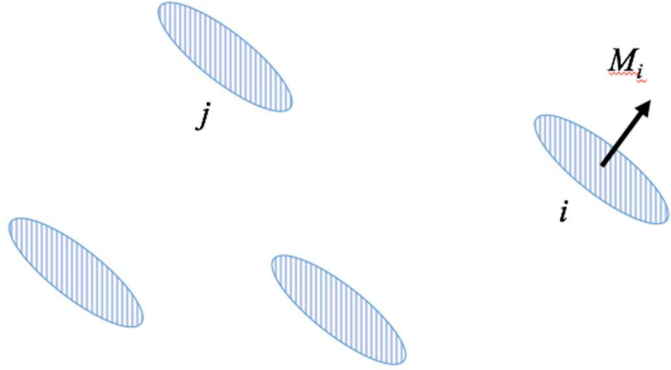

Fig. A1 The shaded regions antiferromagnetically ordered clusters of spins each with magnetization $M_i$.

The free energy of the system as a function of the magnetization $M_i$ of the clusters is given by

$$F = \sum_i g(M_i) - \frac{1}{2}\sum_{i,j} J_{ij} M_i M_j, \qquad (A\ 1)$$

where the indices $i$ and $j$ label the clusters; $g(M)$ is a large non-negative function (the same for all sites $i$) with a minimum at $M = 0$; the coupling constants $J_{ij} = J(\mathbf{r}_i - \mathbf{r}_j)$ are weak. We note that the functions $g(M)$ and $J(\mathbf{r})$ in general depend on temperature. The coupling $J_{ij}$ is assumed to have zero average with respect to averaging over the locations $\mathbf{r}_i$ and $\mathbf{r}_j$ of the clusters, $\int J_{ij} d\mathbf{r}_i\, d\mathbf{r}_j = 0$, as is the case for dipole-dipole interactions or effective interactions in frustrated magnetic systems with quenched disorder. The exact form of the interaction function is not important for the analysis of this section. At the end of the section, however, we consider the special case of dipole-dipole interactions. The function $g(M)$ ensures that the thermodynamic average of the magnetization at each site $i$ vanishes, $\langle M_i \rangle = 0$. Because all the site magnetizations fluctuate along one direction, they commute with the Hamiltonian of the system and may be considered, without loss of generality, as classical (discrete) variables.

In order to perform the averaging of the properties of the system with respect to the locations of the spin clusters, we employ the replica trick. The replicated partition function of the system before the averaging is given by

$$Z^n = Tr_M \left\{ \exp\left(-\frac{1}{T}\left[\sum_{i,\alpha} g(M_i^\alpha) - \frac{1}{2}\sum_{i,j,\alpha} J_{ij} M_i^\alpha M_j^\alpha\right]\right)\right\}, \tag{A 2}$$

where the trace $Tr_M\{...\}$ is taken over all quantum states of the magnetization, $\alpha$ is the replica index and $n$ is the number of replicas. Quenched disorder in the system is represented by the couplings $J_{ij} = J(\vec{r}_i - \vec{r}_j)$ between the randomly located spin clusters. Averaging the partition function over the locations $\vec{r}_i$ of the clusters generates terms of higher orders in the magnetizations $M_i^\alpha$ in the action of the partition function (A 2):

$$\langle \exp\left(\frac{1}{2T}\sum_{i,j,\alpha} J_{ij} M_i^\alpha M_j^\alpha\right)\rangle_{dis} = \exp\left[\frac{1}{2T}\sum_{i,j,\alpha}\langle J_{ij}\rangle_{dis} M_i^\alpha M_j^\alpha + \right. \tag{A 3}$$

$$\left. \frac{1}{8T^2}\sum_{i,j,k,l,\alpha,\gamma}(\langle J_{ij}J_{kl}\rangle_{dis} - \langle J_{ij}\rangle_{dis}\langle J_{kl}\rangle_{dis}) M_i^\alpha M_j^\alpha M_k^\gamma M_l^\gamma + \cdots \right],$$

where $\langle ... \rangle_{dis} = \int \frac{d\vec{r}_1}{V} ... \int \frac{\vec{r}_N}{V} ...$ indicates the averaging.

Below, we develop a mean-field theory of the glass transition, where the correlators of the magnetization between different replicas serve as the order parameter, similarly to conventional models of glass transitions [1,2]. In the mean-field approximation, the order parameter fluctuates weakly around its average value near the transition. Such an approximation allows us to neglect the cumulants above second order (in the couplings $J_{ij}$) in the expansion in Eq. (A 3) near the transition. Going beyond the mean-field approximation or taking into account the higher-order cumulants will lead to corrections of order unity to the temperature of the glass transition and will not affect our results qualitatively.

Since $\langle J_{ij}\rangle_{dis} = 0$ for the interactions under consideration, the first term in the cumulant expansion (A 3) vanishes and the averaging generates quartic interactions. The second cumulant does not vanish if $i = k, j = l$ or $i = l, j = k$. Introducing the notation

$$J_2 = \int (J(r))^2 dr \tag{A 4}$$

and utilizing that $\langle J_{ij}^2 \rangle = J_2/V$, we arrive at the disorder-averaged partition function $\langle Z^n \rangle_{dis} = \int \mathcal{D}M \, exp(-f(M)/T)$ with the effective free energy

$$f(M) = \Sigma_{i,\alpha} g(M_i^\alpha) - \frac{J_2}{4VT} \Sigma_{i,j \neq i,\alpha,\gamma} M_i^\alpha M_i^\gamma M_j^\alpha M_j^\gamma. \tag{A 5}$$

Decoupling the correlators $\frac{1}{V}\Sigma_i M_i^\alpha M_i^\gamma$ for $\alpha \neq \gamma$ in the quartic term by the Hubbard-Stratonovich field $Q^{\alpha\gamma}$, we arrive at the disorder-averaged partition function

$$\langle Z^n \rangle_{dis} = Tr_M \left\{ \int \mathcal{D}Q \, exp \left( -\frac{1}{T}\Sigma_{i,\alpha} g(M_i^\alpha) + \frac{1}{4VT^2}\Sigma_{i,\alpha,\gamma} J_2 \left(M_i^\alpha M_i^\gamma\right)^2 - \right. \right. \tag{A 6}$$
$$\frac{J_2}{4VT^2}\Sigma_{i,j \neq i,\alpha}(M_i^\alpha)^2 \left(M_j^\alpha\right)^2 + \frac{J_2}{2T^2}\Sigma_{i,\alpha,\gamma \neq \alpha} M_i^\alpha M_i^\gamma Q^{\alpha\gamma} -$$
$$\left. \left. \frac{VJ_2}{4T^2}\Sigma_{\alpha,\gamma \neq \alpha}(Q^{\alpha\gamma})^2 \right) \right\}.$$

The second term in the exponent in Eq. (A 6) may be neglected in the thermodynamic limit $V \to \infty$. The role of the third term is a small renormalization of the fluctuations of the magnetizations. The glass transition is associated with the replica-off-diagonal correlators quantified by the order parameters $Q^{\alpha\gamma}$.

Integrating out the magnetizations $M_i^\alpha$ results in the effective free energy in terms of the fields $Q^{\alpha\gamma}$ given by

$$f(Q) = f_0 + \frac{V}{4T}\Sigma_{\alpha,\gamma \neq \alpha} J_2 \left(Q^{\alpha\gamma}\right)^2 - \frac{J_2^2}{8T^3} V n_0 \Sigma_{i,\alpha,\gamma \neq \alpha}[(Q^{\alpha\gamma})^2 + \tag{A 7}$$
$$Q^{\alpha\gamma} Q^{\gamma\alpha}]\langle M^2 \rangle_0^2 + \cdots,$$

where $f_0$ is a constant independent of $Q^{\alpha\gamma}$ and $\langle \ldots \rangle_0 = Tr_M \left\{ \ldots exp\left[-\frac{1}{T}\Sigma_{i,\alpha} g(M_i^\alpha) - \frac{1}{4VT^2}\Sigma_{i,j \neq i,\alpha} J_2(M_i^\alpha)^2 \left(M_j^\alpha\right)^2\right] \right\} / Tr_M \left\{ exp\left[-\frac{1}{T}\Sigma_{i,\alpha} g(M_i^\alpha) - \frac{1}{4VT^2}\Sigma_{i,j \neq i,\alpha} J_2(M_i^\alpha)^2 \left(M_j^\alpha\right)^2\right] \right\}$ is the averaging over the states of isolated spin clusters, in the absence of coupling between them; $Vn_0$ is the number of clusters and ... are the terms of higher orders in $Q$; $\langle M^2 \rangle_0$ is the average of the square of the magnetization in the absence of the glass order parameters.

Here, we look for the replica-symmetric solution $Q_{\alpha\gamma}=q$ for $\alpha \neq \gamma$. In general, the glass phase is replica-symmetry-breaking [1,2]. However, using the replica-symmetric approximation results in a correct order-of-magnitude estimate of the phase boundary. Under the replica-symmetric approximation, the replica free energy is given by

$$f(q) = q^2 \cdot n(n-1)\frac{VJ_2}{4T}\left[1 - \frac{J_2 n_0}{T^2}\langle M^2\rangle^2\right] + O(q^4), \qquad (A\ 8)$$

where n is the number of replicas; the quantity $J_2$ is given by Eq. (A 4) and we have omitted the terms independent of $q$.

The free energy near the transition (i.e. in the limit of the vanishing order parameter $q$) is given by

$$F(q) = \lim_{n\to 0}\frac{f(q)}{n} = q^2 \frac{VJ_2}{4T}\left[\frac{J_2 n_0}{T^2}\langle M^2\rangle^2 - 1\right] + O(q^4) \qquad (A\ 9)$$

and signals a glass phase transition at

$$\frac{J_2 n_0}{T^2}\langle M^2\rangle^2 = 1. \qquad (A\ 10)$$

We emphasize that $J_2$ and $\langle M^2\rangle$ in general depend on temperature.

Equation (A 8) is our main theoretical result for the phase boundary of the glass phase. For the case of fluctuations parallel to one direction considered here, their variance is related to the magnetic susceptibility $\chi(T)$ of the system of the spin clusters by the fluctuation-dissipation theorem as

$$\chi(T) = \frac{n_0\langle M^2\rangle}{T}, \qquad (A\ 11)$$

and the condition for the transition turns into

$$1 = \frac{\chi^2}{n_0}\int [J(\mathbf{r})]^2 d\vec{r}. \qquad (A\ 12)$$

*The form of the interaction function.* Realistic interaction functions $J(r)$ decay with distance $r$ faster than $1/r^{3/2}$, which is why the integral in Eq. *(A 14)* is dominated by the shortest characteristic length scale $r_0$ of the inter-cluster coupling. For example, in the case of dipole-dipole interactions between the clusters, $J(\vec{r}) \propto 1/r^3$, the scale $r_0$ is given by the size of the clusters and plays the role of the short-distance cutoff of the interactions. The quantity $J_2$ is, therefore, given by

$$J_2 \sim 1/(\mu^2 r_0^3), \qquad (A\,13)$$

where $\mu$ is a dimensionless constant (which in the case of dipole-dipole interactions is given by the magnetic permeability of the cloud spins), which results in the condition for the glass transition in the form

$$\mu^2 n_0 r_0^3 \sim \chi^2. \qquad (A\,14)$$